\documentclass[fleqn,usenatbib]{rasti}
\usepackage{newtxtext,newtxmath}
\usepackage[T1]{fontenc}
\usepackage{hyperref}
\usepackage{array}
\usepackage{geometry}
\usepackage{tabularray}
\usepackage{multicol}
\usepackage{booktabs}
\usepackage{tablefootnote}
\usepackage{fancyvrb}
\usepackage{graphicx}	
\usepackage{amsmath}

\usepackage{amssymb}
\usepackage{tikz}
\usetikzlibrary{positioning, quotes}
\usepackage{subcaption}

\hypersetup{linkcolor={blue}, citecolor={blue}, urlcolor={blue}} 
\newcommand{\NAME}{STILTS-NLI}
\newcommand{\new}{} 

\title{\NAME: A Natural Language Interface for STILTS}
\title[STILTS-NLI]{\NAME: A Natural Language Interface for STILTS}
\author{
R. A. Shaw,
S. Fotopoulou,
M. Taylor,
M. Bremer
}
\author[R. A. Shaw et al.]{
R. A. Shaw,$^{1}$\thanks{E-mail: rhys.shaw@bristol.ac.uk}
S. Fotopoulou,$^{1}$
M. Taylor$^{1}$
and M. Bremer$^{1}$
\\
$^{1}$School of Physics, HH Wills Physics Laboratory, Tyndall Ave, Bristol BS8 1TL\\
}

\date{Accepted XXX. Received YYY; in original form ZZZ}

\pubyear{2025}

\begin{document}
\label{firstpage}
\pagerange{\pageref{firstpage}--\pageref{lastpage}}
\maketitle

\begin{abstract}
    The Starlink Tables Infrastructure Library Tool Set (STILTS) is a powerful suite for astronomical data analysis, particularly useful when dealing with large datasets. However, like other software suites in astronomy its comprehensive syntax creates a significant learning curve to new users. To address this, we present \NAME, a natural language interface that generates STILTS commands from user prompts, with agentic support for a user-friendly experience. We developed \NAME\ by fine-tuning a compact, open-source Large Language Model (LLM) on a synthetically generated dataset. This dataset was curated and validated to ensure both comprehensive coverage of key STILTS functionalities and the syntactic correctness of the resulting commands. Our results demonstrate that this specialised model generates valid commands that match and in some cases outperform larger proprietary models. By leveraging small, open-source models, \NAME\ provides an accessible, low-resource solution that lowers the barrier to entry for using STILTS.
\end{abstract}


\section{Introduction}
Starlink Tables Infrastructure Library Tool\footnote{\url{https://www.star.bris.ac.uk/~mbt/stilts/}} \cite[STILTS;][]{taylor_stilts_2006} provides a powerful and versatile suite of command-line tools for processing tabular data, particularly large and complex data sets that are common in astronomy. 
STILTS is complementary to Tool for OPerations on Catalogues And Tables\footnote{\url{https://www.star.bris.ac.uk/~mbt/topcat/}} \cite[TOPCAT;][]{taylor_topcat_2005}. 
TOPCAT is an interactive graphical viewer that provides much of the same functionality as STILTS with a focus on data exploration. 
Both of these tools are very popular among astronomers and are being actively developed.  
STILTS has a large range of functions and parameters to support a wide range of operations on tabular data such as source catalogues. 
Interacting with a set of functions through the command line is common in pipeline processing within the astronomy community.  
STILTS, compared to TOPCAT, is rather under-used and some of this can be attributed to its initial difficulty in crafting commands.
This problem is not isolated to STILTS, it is a common issue with technical astronomy software.
In this respect STILTS is not unusual compared to other software as a sufficiently complex syntax creates a leaning curve for new users and restricts instant usability.

STILTS is a useful but underused package and is an exemplar software system whose accessibility could be improved. 
We address this accessibility challenge as a proof-of-concept for other astronomy software whilst providing a useful more accessible STILTS interface.
To make this possible we turn to large language models (LLMs).
LLMs are built on deep learning architectures such as the Transformer \cite{vaswani_attention_2017}. 
By training on large amounts of text, LLMs can perform a wide array of natural language tasks, including translation, summarization, and question answering \citep{raiaan_review_2024}.

Through an unsupervised training process on large amounts of text, LLMs build a statistical model that represents the probability that a given sequence of words follows another. 
The underlying Transformer architecture is crucial to this process, using self-attention mechanisms to weigh the importance of different words in the input context, allowing the model to capture intricate relationships and long-range dependencies \citep{sajun_historical_2024}.
The practical result is the ability to generate coherent and contextually relevant text by repeatedly predicting the most plausible next word. 
This general capability can then be honed through fine-tuning, where the model learns to approximate a more specific function, such as the mapping from natural language description to a precise line of code. 

In this work, we present the STILTS Natural Language Interface (STILTS-NLI)\footnote{\url{https://github.com/RhysAlfShaw/stilts-nli}} as a proof-of-concept created by fine-tuning an open-source language model to generate STILTS commands from a task description. 
The current model's capabilities allow the creation of a range of STILTS commands that will be useful to new and adept users of STILTS.
This model is made available as a command-line tool which can be run on accessible hardware.

We limited ourselves to using only open-source methods with small LLMs to ensure that access to the model and required compute would not be a barrier to using the tool, as the model can run on local machines, where there is no reliance on proprietary APIs or unrealistic computational demands.
This allows astronomers to control both the capabilities and invariance of the trained model, as the interface's performance cannot be determined by daily updates to larger general purpose LLMs that will not necessarily improve its ability to generate STILTS commands. This allows the astronomy community to have full control of the tool.

Section \ref{sec:methods} describes the process we used to generate training data and fine-tune the model. We look at the performance of a small 2 billion parameter model and a larger 27 billion parameter model of the same family of models.
In Section \ref{sec:model_evaluation} we showcase the similarity between expected and generated STILTS commands in our evaluation data and compare generated commands for the same prompts with other large-LLM providers.
In Section \ref{sec:software_implementation} we present the the software implementation of STILTS-NLI.
In Section \ref{sec:discussion} we discuss the results, limitations, further work needed to expand the tool's capabilities, and implications this kind of solution might have for accessibility in astronomy.


\section{Methodology}
\label{sec:methods}
\subsection{Synthetic Prompt-Response Pair Generation}
\label{subsec:data}
Creating high-quality training data is a critical step in any machine learning project, directly impacting the utility and accuracy of the final model. 
STILTS is very well documented, providing a valuable starting point for data collection. 
Our strategy for training involved creating prompt and response pairs in `JSON' format that we can use to train our model. 
This requires good and realistic natural language task descriptions of valid commands that can be implemented in STILTS. The resulting model will strive to learn this translation from task description to command.

\subsubsection{STILTS documentation}
STILTS benefits from an extensive and detailed description of each command, argument, and flag available at the user's disposal. However, since the logic of the syntax is easily abstracted, it simply suffices for human-oriented documentation to describe the functionality of each argument. On the other hand, LLMs learn by example, therefore an extensive compilation of explicit examples is necessary.
To generate an initial training dataset, we began by extracting the example commands provided within the official STILTS documentation\footnote{\url{https://www.star.bristol.ac.uk/mbt/stilts/sun256/}}. 
This process yielded a modest 150 training examples. Given the breadth of STILTS's functionality, this small number represents only a few examples per task, failing to cover the extensive range of available parameters in an explicit manner, suitable for fine-tuning language models. For a model to learn effectively from prompt-response pairs, a significantly larger dataset is required, encompassing a wide variety of explicit examples with varying parameter combinations for each task.

To make this initial proof-of-concept tractable, we narrowed our focus from the 49 available tasks to a subset of 12 \new{most useful and} frequently utilised tasks \new{to reduce the time required to generate synthetic data}. These are: \texttt{tpipe}, \texttt{tmatchn}, \texttt{tmatch}, \texttt{tcatn}, \texttt{tcat}, \texttt{tcopy}, \texttt{tapquery}, \texttt{cone}, \texttt{mocshape}, \texttt{pixfoot}, \texttt{plot2plane}, \texttt{plot2sky}. 
Out of these tasks, \texttt{tpipe} is the most general and powerful option while \texttt{tcopy} is a \texttt{tpipe} shortcut.

\subsubsection{Synthetic training data}
\begin{figure}
    \centering
    \begin{tikzpicture}[
        every node/.style = {text=black},
        every edge/.style = {draw, ->},
        every edge quotes/.style = {auto, font=\ttfamily,
          text=black, fill=none, sloped}]
        \node (prompt)[draw, rectangle, align=center, text width=2.5cm] {Stilts Task Curated Prompt};
        \node (AI) [draw, rectangle, align=center, right=of prompt, text width=3cm]  {LLM Generates List of prompt and response pairs};
        \node (eval) [draw, rectangle, align=center, below=of AI, text width=2.5cm] {Evaluate prompts for syntax validity};
        \node (accept) [draw, rectangle, align=center, below=1cm of eval, text width=2.5cm]  {Accepted into training data};
        \node (human) [draw, rectangle, align=center, left=1.5cm of eval, text width=2.5cm]  {Manual inspect of failure};
        \node (reject)[draw, rectangle,align=center, below=of human,text width=2cm]  {Reject pair};
        
        \draw (prompt) edge[""] (AI);
        \draw (AI) edge[""] (eval);
        \draw (eval) edge["Valid", rotate=90] (accept);
        \draw (eval) edge["Invalid"] (human);
        \draw (human) edge["Amended"] (accept);
        \draw (human) edge["Rejected", rotate=90] (reject);
    \end{tikzpicture}

    \caption{Workflow for creating synthetic training data.}
    \label{diagram:syn_data_workflow}
\end{figure}
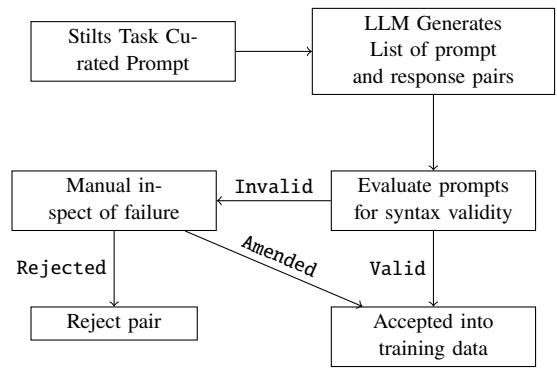

We experimented with several LLMs to generate the synthetic data, aiming for a total of a few thousand prompt-response pairs in JSON format, suitable for model training.
To complement the limited number of examples from the documentation, we developed a workflow to generate and validate synthetic training data, illustrated in Figure \ref{diagram:syn_data_workflow}. 
This process begins by providing a powerful, closed-source LLM with STILTS specific context, in two distinct strategies.

First, we provided Gemini 2.5 Pro, GPT-4.5 and 5, and Claude 3.5, 4.0, 4.5 the documentation of only a specific task each time, including its parameters, existing examples, and special operations. This is due to each model's limited context length, which prevents it from holding the entire STILTS documentation in memory. The second strategy was to provide the single-page HTML STILTS documentation as source in NotebookLM. In this case, the service maps the input source into the model latent space and allows for a much larger context window. Therefore, we were able to ask examples for each command without breaking down the documentation into bite-sized pieces.

In all cases, the models were instructed to generate new, diverse examples. These instructions guide the model to vary the language of the prompts, target different data manipulation goals, and utilise a wide array of each task's arguments. We were explicit to mention that we need a variety of examples for fune-tuning a language model.

We noticed a difference in the strategy of each model to complete the requested task. Models available before August 2025, showed a higher degree of hallucination in the command generation, particularly in providing regular expressions that are invalid in STILTS, and especially generating invalid expressions with `NULL'\footnote{The correct syntax to check for null values in a column, e.g. MAG, is \texttt{NULL\_MAG}, while the models often suggested \texttt{isNULL(MAG)}, perfectly reasonable but not correct.}
Models available after August 2025, when requested\footnote{We accessed the models through \texttt{Cursor}.} to generate more than 20 examples at a time, opted to create a python script that cycled through the commands arguments in order to provide good coverage. This led to a very limited number of prompt expressions that lacked a human level of natural expression when describing a task. 

On the other hand, NotebookLM was able to generate about 100 examples with every request and repeating the request to ``please generate 100 more examples'' was enough to gather the training data after subsequent prompting. We found NotebookLM commands to be more natural and diverse in the prompt description and generated more accurate STILTS commands, therefore we adopted this training set for the remainder of this work. Anecdotally, even though we requested an exact number of prompt-response pairs every time, the model instead generated approximately the correct order of magnitude. Hence, our training set covers about 1,000 examples of the commands \texttt{tpipe} and \texttt{tapquery}, and about 500 examples for each of the other commands. We also noticed occasional repetition of the past answers instead of generating a new set, and few minor typos on the \texttt{JSON} formatting, such as very occasionally missing the word ``prompt''.

\subsubsection{Validation of training data}
Finally, each generated command undergoes a validation process.
We first perform an automated syntactic check using a call to an internal STILTS function (which is not directly accessible), which validates the syntax of STILTS tasks and arguments. Unfortunately, the validity of functions within \texttt{cmd} arguments are not validated in this function call, so additional checks were necessary to ensure that no hallucinated functions were present in the training data.
Commands that fail this check are manually inspected to determine if minor corrections can be made; otherwise, they are discarded.
This two-step validation process ensures that our final training dataset consists exclusively of syntactically valid STILTS commands.
This resulted in 7,270 raw prompt-response pairs. This is summarised in Table \ref{tab:data_summary} and broken down by STILTS task. 
After the checks described above, the training dataset was reduced to 6,228 pairs equating to losing $\approx15\%$ of generated pairs. This corresponds to about 1,000 examples removed from the training set, which is significantly more than the 250 bad examples needed to poison a model \cite{souly_poisoning_2025}.

Three prompt-response examples are shown in Table \ref{tab:finetuning-examples}. As more complicated commands are inherently difficult to create, we expected these to be more likely to be rejected by these checks, reflected in the higher invalid rate of the more flexible commands such as \texttt{tpipe} and \texttt{tmatchn}.

Fixes to failed syntax checked pairs were made possible due to common failure modes. 
The most prominent of these, present mostly in \texttt{tpipe}, was due to missing quotes inside operations e.g. \texttt{cmd='select MAG\_u > 10'} should be \texttt{cmd='select "MAG\_u > 10"'}. 
This consistent failure allowed for them to be fixed through in part to using a regular expression\footnote{\texttt{((?:o\textbar i\textbar u)?cmd\textbackslash d*)=\textbackslash x27([\^{}\textbackslash x27]*)\textbackslash x27} - This finds strings beginning with \texttt{cmd}, which can be prefixed by an 'o', 'i', or 'u'. The regular expression then captures the value enclosed within a pair of single quotes that follows \texttt{cmd=}.} to find all these instances and add the necessary quotes. 
Within \texttt{cone} pairs a common failure was due to a providing a mathematical expression as input which STILTS does not evaluate before execution, e.g. \texttt{radius='20/60'} should be \texttt{radius=0.33}. 
This is present when the radius is given as arcminutes or arcseconds and the function here is the unit conversion to degrees. 
Since these are not technically incorrect we decided to keep them in the dataset, but draw attention to this as a expected behaviour and the user of this model will have to edit this manually. Without introducing explicit examples of varying angular units we found that the model would always use the number requested by the user, regardless of the expected unit in each STILTS command.

Finally, we faced the challenge of validating commands that expect to read a file with the correct column names in order to automatically check the validity of the command, which was not currently feasible to produce while using a wide variety of file and column names. These were the majority of the removed \texttt{tpipe} pairs which contained parameters for the task \texttt{tgroup} that handles table aggregation. We decided not to support this functionality.
Similarly, we were unable to validate plotting tasks \texttt{plot2plane} and \texttt{plot2sky}. 
We decided to include these tasks that failed on this basis as the commands look convincing, but we advise that these tasks should be used with caution and treated as a first attempt to craft a successful command. \new{Even though verifying the semantic validity of the synthetic prompts was not done systematically, we did visually inspect the training data. Although prompts appeared to be appropriate for the specific command, the lack of rigorous validation remains a limitation of our method.}

\begin{table}
\centering
\begin{tabular}{>{\centering\hspace{0pt}}m{0.17\linewidth}|>{\centering\hspace{0pt}}m{0.07\linewidth}>{\centering\hspace{0pt}}m{0.096\linewidth}>{\centering\hspace{0pt}}m{0.129\linewidth}>{\centering\hspace{0pt}}m{0.078\linewidth}>{\centering\hspace{0pt}}m{0.071\linewidth}>{\centering\arraybackslash\hspace{0pt}}m{0.061\linewidth}} 
\hline\hline
Task&Total&Invalid&Removed&Fixed&Final&Valid\par{}(\%)  \\ 
\hline\hline
cone       & 516   & 291     & 77      & 215   & 440   & 85.3             \\
mocshape   & 533   & 206     & 163     & 78    & 405   & 76.0             \\
pixfoot    & 517   & 217     & 172     & 75    & 375   & 72.5             \\
plot2plane*& 524   & 489     & 65      & 443   & 478   & 91.2             \\
plot2sky*  & 513   & 508     & 31      & 478   & 483   & 94.1             \\
tapquery   & 1040  & 100     & 41      & 59    & 999   & 96.1             \\
tcat       & 539   & 195     & 136     & 83    & 427   & 79.2             \\
tcatn      & 537   & 134     & 60      & 84    & 487   & 90.7             \\
tcopy      & 507   & 51      & 58      & 1     & 457   & 90.1             \\
tmatch2    & 504   & 92      & 49      & 55    & 467   & 92.7             \\
tmatchn    & 500   & 154     & 148     & 41    & 387   & 77.4             \\
tpipe      & 1040  & 619     & 255     & 402   & 823   & 79.1             \\ 
\hline
Total       & 7270  & 3056    & 1042    & 2014  & 6228  & 85.7             \\
\hline
\end{tabular}
\small
* Plotting task failures due to missing files were accepted.
\caption{Summary table of STILTS training data, generated, removed, fixed and invalid. This is a total of 7270 generated prompt-response pairs reducing to 6228 pairs for training.}
\label{tab:data_summary}
\end{table}

\begin{table*}
    \centering
    \begin{tabular}{p{0.45\linewidth} p{0.45\linewidth}}
        \textbf{Prompt} & \textbf{Response} \\
        \toprule
        \textit{From `input.fits`, select all sources within a 15 arcminute radius of RA=150.25, Dec=2.5. The table's coordinate columns are 'RA\_DEGREES' and 'DEC\_DEGREES'. Count how many matching sources are found.} & \texttt{stilts tpipe in=input.fits cmd='select "skyDistanceDegrees(RA\_DEGREES, DEC\_DEGREES, 150.25, 2.5) < 15./60."' omode=count} \\
        \\
        \textit{From each input table ('input1.fits', 'input2.fits'), I only want to keep the columns 'RA', 'DEC', and 'PM\_RA'. Concatenate these filtered tables and save to 'output.fits'} & \texttt{stilts tcat in='input1.fits input2.fits' icmd='keepcols "RA DEC PM\_RA"' out=output.fits} \\     
        \\
        \textit{Perform an integrity check on the table `input.fits` and report any issues found to the console.} & \texttt{stilts tpipe in=input.fits cmd='check'} \\
        
        \bottomrule
    \end{tabular}
    \caption{Example synthetic Prompt-Response pairs created by the process described in Figure \ref{diagram:syn_data_workflow}. }
    \label{tab:finetuning-examples}
\end{table*}


\subsection{Training}
\label{subsec:Training}

The training process for our model was conducted in two distinct phases: a domain-adaptive pre-training phase to familiarise the model with the specific syntax and vocabulary of the STILTS software, followed by a fine-tuning phase to teach the model the task of translating prompts into executable commands. 
We used Google's Gemma 2 billion\footnote{\url{https://huggingface.co/google/gemma-2b}} (2B) and 27 billion\footnote{\url{https://huggingface.co/google/gemma-2-27b}} (27B) parameter models \citep{gemma_team_gemma_2025} as a foundation, accessed through HuggingFace. \new{This model family was chosen because of its compact size and the performance it showed in some preliminary experiments.}

Training was implemented using the \textit{transformers} library \citep{wolf_transformers_2020}, a standard framework for training transformer-based models, with a single NVIDIA A6000 GPU for the 2B parameter model. For the larger 27B parameter model we used one compute node on Isambard-AI using four NVIDIA GH200 Grace Hopper Superchips \citep{mcintosh-smith_isambard-ai_2024}. All code for model training is available on GitHub\footnote{\url{https://github.com/RhysAlfShaw/stilts-nli-train}}.
\new{For both the pre-training and fine-tuning steps, we used standard cross-entropy loss to determine the difference between the desired and output probability distributions.}

\subsubsection{Pre-training}
\label{subsubsec:Pretraining}
The primary goal of the pre-training phase was to imbue a general-purpose foundation model with specialised knowledge of the STILTS ecosystem and create embeddings for the vocabulary used in STILTS.
To achieve this, we extracted and cleaned the full STILTS documentation \new{by removing HTML and other unnecessary tokens from the the documentation source}. This \new{documentation} includes detailed descriptions of tasks, lists of parameters and their details along with command examples and use cases. 
These raw textual data were then chunked to 1024 tokens for training. Both models were trained for only a single epoch, i.e. they were exposed to the entire documentation only once. Pre-training was not long and only took $\sim2$ and $\sim5$ minutes for the 2B and 27B models respectively.

\subsubsection{Fine-tuning}
\label{subsubsec:fine-tunning}
The synthetic prompt-response dataset, generated as described in section \ref{subsec:data}, was used to fine-tune our pretrained model. The evaluation dataset was selected as a random 10\% of the data generated and validated for training.
We used the AdamW optimiser \citep{kingma_adam_2017} with an initial learning rate of 
$5\times10^{-5}$ and $1\times10^{-5}$ for 2B and 27B models, respectively, with a batch size of 2. This took $\sim$30 minutes for the 2B parameter model and $\sim$12 hours for the 27B model.

\subsection{Model Evaluation}
\label{sec:model_evaluation}

To evaluate our trained model we examined its performance on the 10\% of the training data not used for training (622 prompt-response pairs). To measure how similar the generated and expected STILTS commands are we use the cosine similarity. This measures the cosine of the angle ($\theta$) between the two high dimensionality vectors created by an embedding model\footnote{We used \texttt{sentence-transformers/all-MiniLM-L6-v2} model\footnote{\url{https://huggingface.co/sentence-transformers/all-MiniLM-L6-v2}} \citep{reimers_sentence-bert_2019}}. The cosine similarity for two vectors $A$ and $B$ is expressed as the following: 
\begin{equation}
\text{similarity} = \cos(\theta) = \frac{A \cdot B}{\|A\| \|B\|} = \frac{\sum_{i=1}^{n} A_i B_i}{\sqrt{\sum_{i=1}^{n} A_i^2} \sqrt{\sum_{i=1}^{n} B_i^2}}.
\end{equation}
A score of 1 indicates an identical response and maximum semantic similarity, 0 indicates orthogonality meaning no semantic similarity. 

This is not an infallible method for evaluation as a perfectly matching generated command might not be able to be created given the information in the prompt. 
For example, this is shown in some of these data as prompts that do not specify a specific input or output file and/or column names usually contain different yet reasonable file and column names. The total cosine similarity for the evaluation data set is shown in Figure \ref{fig:overal_sim_histogram} and the similarity categorised by task can be found in appendix \ref{fig:sim_hist_by_task}, the evaluation data has a mean similarity of $0.91\pm0.11$ and $0.89\pm0.12$ for the 2B and 27B models respectively. As the smaller 2B parameter model and much larger 27B parameter model have comparable performance across every task we only make the 2B model available for use with STILTS-NLI.
\new{We are not assessing the correctness of the function that the command creates, this score tells us how well the trained model performs at generating a prompt from the same distribution as the training data. For an assessment on the functional assessment of the trained model see section \ref{sec:comp_to_gen_models}}.

\begin{figure}
    \centering
    \includegraphics[width=\linewidth]{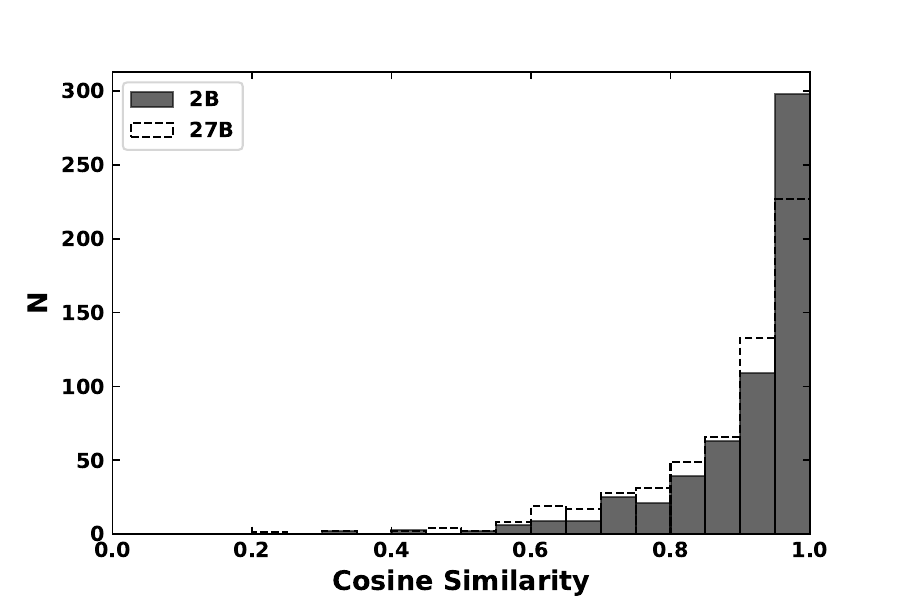}
    \caption{Cosine similarity distribution of the evaluation data. For both the 2B (shaded grey) model and the larger 27B model (dashed outline).}
    \label{fig:overal_sim_histogram}
\end{figure}

\begin{figure}
    \centering
    \includegraphics[width=\linewidth]{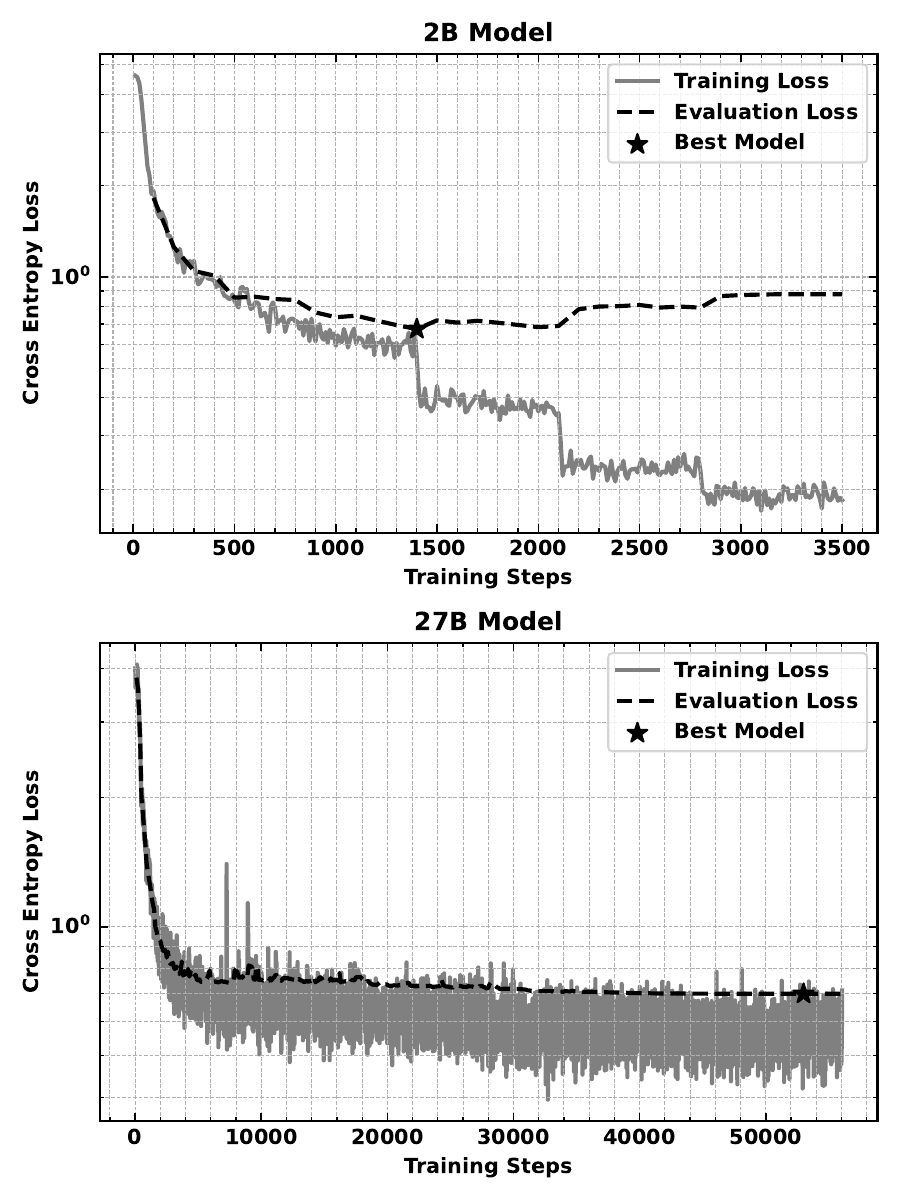}
    \caption{Training and evaluation loss curves from fine-tuning the 2B (top) and 27B (bottom) models. }
    \label{fig:loss_curve}
\end{figure}

The training and evaluation loss curves can be seen in Figure \ref{fig:loss_curve}. 
We see that the 2B and 27B models perform best on the unseen evaluation set at 1,400 and 52,500 training steps respectively. 
At later steps, we notice that the 2B model begins to overfit to the training data and we see sudden decreases in loss which occur periodically at every epoch. The number of training steps is calculated as,
\begin{equation}
N_{steps} = \frac{N_{data}/S_{batch}}{N_{Grad}},
\end{equation}
where $N_{Grad}$=4, $S_{batch}$=2 and $N_{data}$=5606. $N_{Grad}$ tells the model how many gradients from different batches of size $S_{batch}$ to accumulate before preforming a single update to the model. 
From this we expect a new epoch to start every 701 training steps. These features are present at multiples of 701 so we know they are happening as the model see the same training data again.
The 27B parameter model takes much longer to train both in terms of time per training step and the number of steps required. 
This is because of the model's drastically larger number of weights. 
In both cases, the weights that minimise the evaluation loss were adopted as the final trained model, this is shown in Figure \ref{fig:loss_curve} with a star.


\section{Agentic Software Implementation}
\label{sec:software_implementation}

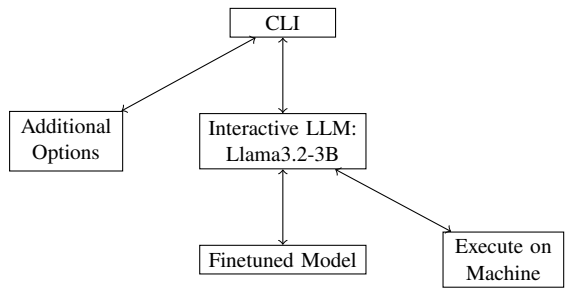
\begin{figure}
    \centering

    \begin{tikzpicture}[
        every node/.style = {text=black},
        every edge/.style = {draw, <->},
        every edge quotes/.style = {auto, font=\ttfamily,
          text=black, fill=none, sloped}]
        \node (CLI) [draw, rectangle,align=center, text width=1.2cm] {CLI};
        \node (llm-interactive) [draw, rectangle,align=center, below=of CLI, text width=2cm] {Interactive LLM: Llama3.2-3B};
        \node (Finetuned-model) [draw, rectangle,align=center, below=of llm-interactive, text width=2cm] {Finetuned Model};
        \node (additional options) [draw, rectangle,align=center, below=of CLI, left=of llm-interactive,text width=1.3cm]{Additional Options};
        \node (run) [draw, rectangle,align=center, below=of llm-interactive, right=of Finetuned-model,text width=1.4cm] {Execute on Machine};
        
        \draw (CLI) edge (llm-interactive);
        \draw (llm-interactive) edge (Finetuned-model);
        \draw (llm-interactive) edge (run);
        \draw (CLI) edge (additional options);

    \end{tikzpicture}
        
    \caption{STILTS-NLI implementation.}
    \label{diagramsoftware_implementation_workflow}
\end{figure}

Fine-tuning an LLM on prompt-response pair results in the model only responding with STILTS commands. This provides a poor user experience as every input will be translated into a command.
To make the interface more user friendly, and allow for continuous command development, we implemented an instruction-trained LLM to act as an intermediary. 
This model can follow more diverse user instruction and needed to be able to take in descriptions of tools from the system prompt that point to other models or external functions.
For this we chose Meta's Llama-3.2-3B-Instruct model\footnote{\url{https://huggingface.co/meta-llama/Llama-3.2-3B-Instruct}} \citep{grattafiori_llama_2024}.
External function calls are enabled by providing a schema of the available functions within the system prompt.
When the model thinks this needs to be called it will only return the function command and fill in any necessary parameters, then the model will return a python-like function.
If this response is found in the model's output we call that external function. In our case we either call the fine-tuned model or attempt to execute the STILTS command on the users machine. This implementation is shown in Figure \ref{diagramsoftware_implementation_workflow}.
In this 'agentic' mode the user does not directly interact with the fine-tuned model but will have its prompts forwarded to it.
To further increase accessibility to this tool the \verb|--stilts_model_only| flag can be parsed on startup to interact directly with the fine-tuned model; this reduces the amount of RAM need by STILTS-NLI from $\sim$12GB to $\sim$5GB.

Additional options to quit (\texttt{q/quit}), save the prompt history (\texttt{s/save}), clear the prompt history (\texttt{c/clear}) and get some guidance (\texttt{h/help}) are available. 
Files within the current directory can be searched for by using a single quote (\verb|'|); this will list local files as the user types and filter the search by matching to the typed characters. This makes it easier for the user to provide exact filenames into STILTS-NLI.

In Appendix \ref{appen:example_user_interaction} we show an example session interacting with \NAME.
Installation instructions for \NAME\ are available on GitHub\footnote{\url{https://github.com/RhysAlfShaw/stilts-nli}} and the fine-tuned models\footnote{\url{https://huggingface.co/RAShaw/gemma-2b-stilts-prototype}} are publicly available on HuggingFace.


\begin{table*}
  \centering
  \setlength{\tabcolsep}{6pt} 
\renewcommand{\arraystretch}{1.5} 
  \begin{tabular}{c|p{90mm}|c|c}
    \hline
    Model & Result & Syntax & Correct \\
    \hline
    \multicolumn{4}{|p{\linewidth}|}{Prompt 1: Create a STILTS command that converts a fits catalogue into a CSV file.} \\ \hline
    \NAME & \texttt{tcopy in=input.fits out=output.csv} & Valid & Yes \\ 
    Gemini 2.5 Pro & \texttt{stilts tpipe in=your\_catalogue.fits out=output\_file.csv ofmt=csv} & Valid & Yes \\
    GPT-5 & \texttt{tcopy in=catalog.fits out=catalog.csv ofmt=csv} & Valid & Yes \\
    \hline
    
    \multicolumn{4}{|p{\linewidth}|}{Prompt 2: Make a new column in catalogue input.fits that is defined as column MAG\_u - MAG\_i and save as a parquet.} \\ \hline
    \NAME & \texttt{tpipe in=catalogue\_input.fits ifmt=fits cmd='addcol MAG\_u\_i "MAG\_u - MAG\_i"' out=catalogue\_with\_color.parquet ofmt=parquet} & Valid & Yes \\
    Gemini 2.5 Pro &\texttt{tpipe in=catalogue\_input.fits cmd='addcol u\_minus\_i "MAG\_u - MAG\_i"' out=catalogue\_output.parquet ofmt=parquet} & Valid & Yes \\
    GPT-5 & \texttt{tpipe in=input.fits cmd='addcol MAG\_u\_minus\_i "MAG\_u - MAG\_i"' out=output.parquet} & Valid & Yes\\
    \hline
    
    \multicolumn{4}{|p{\linewidth}|}{Prompt 3: Create a STILTS command that will change any missing or nan values to -99.} \\ \hline
    \NAME & \texttt{tpipe in=input.fits cmd='replaceval null -99 *' out=output.fits} & Valid & Yes \\
    Gemini 2.5 Pro & \texttt{tpipe in=input.fits out=output.fits cmd='replaceval null -99 *'} & Valid & Yes \\ 
    GPT-5 & 
    \texttt{tpipe in=your\_input\_table.fits cmd='-update x, (isnan(x) ? -99 : x)' out=cleaned\_table.fits}
    & Invalid & No\\
    \hline

    \multicolumn{4}{|p{\linewidth}|}{Prompt 4: Match source\_cat.fits, pybdsf\_cat.fits, sdss\_source.fits, aux\_source.fits using group mode and filter pybdsf\_cat.fits to keep only rows where column ERROR is not null.} \\ \hline
    \NAME & \texttt{tmatchn nin=4 multimode=group matcher=sky params=2 in1=source\_cat.fits in2=pybdsf\_cat.fits icmd2='select !NULL\_ERROR' in3=sdss\_source.fits in4=aux\_source.fits values1='RA DEC' values2='RA DEC' values3='RA DEC' values4='RA DEC' out=clean\_pybdsf\_group.fits} & Valid & Yes \\
    Gemini 2.5 Pro & \texttt{tmatchn nin=4 join=group in1=source\_cat.fits in2=pybdsf\_cat.fits icmd2='select "!isNull(ERROR)"' in3=sdss\_source.fits in4=aux\_source.fits matcher=sky params=1 values1="RA DEC" values2="RA DEC" values3="RA DEC" values4="RA DEC" out=matched\_cat.fits} & Invalid & No \\ 
    GPT-5 & 
    \texttt{tmatchn nin=4 in1=source\_cat.fits in2=pybdsf\_cat.fits in3=sdss\_source.fits in4=aux\_source.fits matcher=sky params=1 join=all group=each ifmt1=fits ifmt2=fits ifmt3=fits ifmt4=fits ocmd2='select "ERROR != null"' out=matched\_group.fits ofmt=fits}
    & Invalid & No \\
    \hline
    
  \end{tabular}
  \caption{Response comparison between our STILTS fine tuned model, Gemini 2.5 Pro and GPT-5 on 3 example prompts with increasing difficulty. All models prompted in September and October 2025.}
  \label{tab:prompt_comparison}
  
\end{table*}

\section{Discussion}
\label{sec:discussion}

\subsection{Comparison to general purpose models}
\label{sec:comp_to_gen_models}
To understand how our model (2B only) compares with premium large-LLM providers, we show results from our model, Gemini 2.5 Pro and GPT-5 (free version).
We evaluate the resulting command using our syntax checker, as defined in Section \ref{subsec:data}, and whether the described task is completed by this result. The results are shown in Table \ref{tab:prompt_comparison}.
We examine the performance of different prompts an astronomer would use with increasing difficulty,
\new{these prompts are not taken from our training or validation data}.
Starting with file conversions, creating colours, changing any missing data to '-99', e.g. typically used when fitting photometry to SEDs, and a four-way catalogue merger with a filter on only one of those catalogues.
Different models here omit or include the output file type argument \texttt{ofmt}, this format is not strictly necessary as STILTS can infer the output file type based on the file extension. 

Our fine-tuned model is able to generate syntactically correct commands for all four test prompts. 
We see that both Gemini and GPT-5 succeed at the more basic prompts, but fail on the more specific and complicated tasks. 
The failures in all cases are due to hallucinated functions within the \texttt{cmd} argument with the addition of an invalid \texttt{join} argument for prompt 4 for both Gemini and GPT-5. This demonstrates a key advantage of using validated, subject-specific datasets: they prevent model hallucinations by ensuring the model is adequately trained on the relevant domain and syntax.
This demonstrates the viability of using a fine-tuned, small, open-source language model to create an accessible natural language interface for STILTS. 
The developed tool, which can be used with everyday performance computers, successfully translates user descriptions into syntactically valid STILTS commands\footnote{We note that the final responsibility for building a precise command always lies with the user, as the model might not have been given enough information in the prompt.}.

A key finding from our comparative analysis is that our specialised, fine-tuned model can match and even outperform much larger, general-purpose, closed-source models like Gemini 2.5 and GPT-5 on domain-specific tasks. 
\new{This result is supported by \cite{tamhane_natural_2025} who also showed that small models can preform well at a similar astronomy-specific query task.}
As shown in Table \ref{tab:prompt_comparison}, while all models handled simple requests adeptly, GPT-5 failed on the more complex prompts. 
GPT-5 was the lowest performer as it hallucinates multiple functions in both prompt 3 and 4, whereas STILTS-NLI model was able to generate valid commands.  
This highlights a crucial advantage of fine-tuning, where the validated dataset and specific domain knowledge limits the risk of the model generating invalid syntax. As in the creation of our training data we were able to remove hallucinated functions like \texttt{isNull}. 

The generalist models, despite their vast knowledge, are approximating from a much broader context and can falter when faced with the niche, rigid syntax of a specialised tool like STILTS, especially when it constitutes a minuscule fraction of its total training set. 
Our model’s performance is a direct result of being trained on a curated, syntactically validated dataset, reinforcing the importance of data quality in fine-tuning applications. 

From the comparison between the trained 2B and 27B models we can see that a large model does not necessarily equate to better performance.
In this case, the larger model took a significantly longer period to train and resulted in comparable or slightly inferior performance. 
This is likely due to the significant number of parameters and insufficient data to achieve good performance.
At least for the command subset we have trained on, this model is more complicated than necessary to learn the STILTS functionality. 
If the number of supported commands is increased and significantly more training data are acquired this might change.

\subsection{Accessibility}
The decision to exclusively use open-source models and methods was central to the project’s ethos. 
By building a tool that does not rely on proprietary APIs or require significant computational resources, we ensure it is maximally accessible to the scientific community and has limited environmental impacts due to the model's very small size.

\new{Alternative techniques for adding knowledge to existing models, like an MCP-based method, for STILTS are currently incompatible with our goal of using small, accessible models. STILTS requires detailed tool calling descriptions and small models have consistently been shown to struggle with such complex tasks \citep{shen_small_2024}. 
Whilst future developments could improve small model tool calling \citep{manduzio_improving_2024}, our current strategy ensures STILTS remains open-access allowing students and researchers to benefit without incurring the financial costs of depending on third-party services.}
Furthermore, the agentic two-model strategy a fine-tuned model for the core translation task and a general instruction-tuned model for the user interface provides a user-friendly experience without compromising the specialised accuracy of the command generation.

In addition, this work illustrates an example of a complex astronomy tool that can be adapted to a non-traditional method of interaction. With the addition of a speech-to-text layer, STILTS-NLI could provide a pathway for people who cannot easily use a keyboard to access expert astronomy tools, further broadening its impact on the wider astronomy community.

\subsection{Future extensions}
Despite the promising results, the model can be further improved in a number of ways. The model’s knowledge is constrained to a small number of STILTS functions; it cannot generate commands for other tasks within the STILTS suite and we could not properly validate functions within \texttt{cmd} arguments or tasks \texttt{plot2plane} and \texttt{plot2sky}. The primary bottleneck to expanding this coverage to the full breadth of STILTS commands is the creation of high-quality training data.
The semi-automated process described in Section \ref{subsec:data} is laborious and not perfect; scaling it across the entirety of STILTS’s functionality would require significant effort.
While generating more data is somewhat straightforward, increasing the quality and correctness of that data will require significant expert time and continuous feedback.
Whilst we mostly used Google's NotebookLM to generate training data, we do not believe one model is definitively superior to others.
As these models are continuously developed, the latest version, provided it shows an increase in quality of command complexity accuracy and prompt language, should be used to produce future versions of \NAME. 

Alternatively, we can elicit the input of the astronomical community by collecting suggested prompt-response pairs from users when our trained model fails to create a suitable command.
This would outsource this data collection to the user base and help improve the translation from realistic human description to a valid and correct STILTS command.
Collecting data this way comes with its own issues, related to trusting the user and whether it is reasonable to get the submitted STILTS command from the description. We encourage users of \NAME\ to submit such examples in the GitHub repository on the issues page using the tag \texttt{suggested-pair}.

As LLM models improve, it is likely that open-source models' inference capabilities will also improve. 
Since this model is currently able to run on a typical laptop, the model size — which defines the memory requirement — is likely to remain the same. 
As new devices are made with more RAM and more support for AI models, this solution will become increasingly even more accessible. 
From the results of this work, we believe that the current limitation in performance is the quality and breadth of the training data and not model size.

\section{Conclusion}
\label{sec:conclusion}
We have successfully developed a natural language interface for STILTS by fine-tuning a small, open-source language model, effectively lowering the accessibility barrier for this powerful astronomical software.
Our central finding is that this specialised model can match and outperform large, general-purpose models like GPT-5 and Gemini, particularly on complex commands where it avoids syntax 'hallucinations'.
By using an open-source, locally runnable model, our tool is maximally accessible, private, and computationally efficient.
The primary limitation of this work is that the model's knowledge is confined to the twelve STILTS tasks used in training. 
Future work will focus on expanding the synthetic training dataset to cover the full functionality of the STILTS suite, further validating this approach for creating accessible interfaces to specialised scientific tools.

\section{Data Availability}
All training data and code can be found on GitHub at \url{https://github.com/RhysAlfShaw/stilts-nli} and \url{https://github.com/RhysAlfShaw/stilts-nli-train}. 
Models used in this work are available on HuggingFace at \url{https://huggingface.co/RAShaw/gemma-2b-stilts-prototype}. 
Any bugs in STILTS-NLI can be reported on its issues page \url{https://github.com/RhysAlfShaw/stilts-nli/issues}.

\section{acknowledgements}
\new{We thank the anonymous reviewers for their constructive feedback on this work.}
This work has benefited from the support of Royal Society Research Grant RGS{\textbackslash}R1\textbackslash231450. 
This work was supported by the UKRI Centre for Doctoral Training in Artificial Intelligence, Machine Learning \& Advanced Computing, funded by grant EP/S023992/1. This work was supported by STFC small award grant ST/Y002032/1. 
The authors acknowledge the use of resources provided by the Isambard-AI National AI Research Resource (AIRR).
Isambard-AI is operated by the University of Bristol and is funded by the UK Government’s Department for Science, Innovation and Technology (DSIT) via UK Research and Innovation; and the Science and Technology Facilities Council [ST/AIRR/I-A-I/1023]. 
\new{The authors acknowledge the use of Google Gemini (2025 version) for generating small code snippets, which were subsequently verified and refined by the authors.}

\section{Conflict of Interest statement}
Authors declare no conflict of interest

\bibliographystyle{mnras}
\bibliography{references}

\begin{thebibliography}{}
\makeatletter
\relax
\def\mn@urlcharsother{\let\do\@makeother \do\$\do\&\do\#\do\^\do\_\do\%\do\~}
\def\mn@doi{\begingroup\mn@urlcharsother \@ifnextchar [ {\mn@doi@} {\mn@doi@[]}}
\def\mn@doi@[#1]#2{\def\@tempa{#1}\ifx\@tempa\@empty \href {http://dx.doi.org/#2} {doi:#2}\else \href {http://dx.doi.org/#2} {#1}\fi \endgroup}
\def\mn@eprint#1#2{\mn@eprint@#1:#2::\@nil}
\def\mn@eprint@arXiv#1{\href {http://arxiv.org/abs/#1} {{\tt arXiv:#1}}}
\def\mn@eprint@dblp#1{\href {http://dblp.uni-trier.de/rec/bibtex/#1.xml} {dblp:#1}}
\def\mn@eprint@#1:#2:#3:#4\@nil{\def\@tempa {#1}\def\@tempb {#2}\def\@tempc {#3}\ifx \@tempc \@empty \let \@tempc \@tempb \let \@tempb \@tempa \fi \ifx \@tempb \@empty \def\@tempb {arXiv}\fi \@ifundefined {mn@eprint@\@tempb}{\@tempb:\@tempc}{\expandafter \expandafter \csname mn@eprint@\@tempb\endcsname \expandafter{\@tempc}}}

\bibitem[\protect\citeauthoryear{{Gemma Team} et~al.,}{{Gemma Team} et~al.}{2025}]{gemma_team_gemma_2025}
{Gemma Team} et~al., 2025, Gemma 3 {Technical} {Report}, \mn@doi{10.48550/arXiv.2503.19786}, \url {https://ui.adsabs.harvard.edu/abs/2025arXiv250319786G}

\bibitem[\protect\citeauthoryear{Grattafiori et~al.,}{Grattafiori et~al.}{2024}]{grattafiori_llama_2024}
Grattafiori A.,  et~al., 2024, The {Llama} 3 {Herd} of {Models}, \mn@doi{10.48550/arXiv.2407.21783}, \url {http://arxiv.org/abs/2407.21783}

\bibitem[\protect\citeauthoryear{Kingma \& Ba}{Kingma \& Ba}{2017}]{kingma_adam_2017}
Kingma D.~P.,  Ba J.,  2017, Adam: {A} {Method} for {Stochastic} {Optimization}, \mn@doi{10.48550/arXiv.1412.6980}, \url {http://arxiv.org/abs/1412.6980}

\bibitem[\protect\citeauthoryear{Manduzio, Galatolo, Cimino, Scilingo  \& Cominelli}{Manduzio et~al.}{2024}]{manduzio_improving_2024}
Manduzio G.~A.,  Galatolo F.~A.,  Cimino M. G. C.~A.,  Scilingo E.~P.,   Cominelli L.,  2024, Improving {Small}-{Scale} {Large} {Language} {Models} {Function} {Calling} for {Reasoning} {Tasks}, \mn@doi{10.48550/arXiv.2410.18890}, \url {http://arxiv.org/abs/2410.18890}

\bibitem[\protect\citeauthoryear{McIntosh-Smith, Alam  \& Woods}{McIntosh-Smith et~al.}{2024}]{mcintosh-smith_isambard-ai_2024}
McIntosh-Smith S.,  Alam S.~R.,   Woods C.,  2024, Isambard-{AI}: a leadership class supercomputer optimised specifically for {Artificial} {Intelligence}, \mn@doi{10.48550/arXiv.2410.11199}, \url {http://arxiv.org/abs/2410.11199}

\bibitem[\protect\citeauthoryear{Raiaan et~al.,}{Raiaan et~al.}{2024}]{raiaan_review_2024}
Raiaan M. A.~K.,  et~al., 2024, \mn@doi [IEEE Access] {10.1109/ACCESS.2024.3365742}, 12, 26839

\bibitem[\protect\citeauthoryear{Reimers \& Gurevych}{Reimers \& Gurevych}{2019}]{reimers_sentence-bert_2019}
Reimers N.,  Gurevych I.,  2019, Sentence-{BERT}: {Sentence} {Embeddings} using {Siamese} {BERT}-{Networks}, \mn@doi{10.48550/arXiv.1908.10084}, \url {http://arxiv.org/abs/1908.10084}

\bibitem[\protect\citeauthoryear{Sajun, Zualkernan  \& Sankalpa}{Sajun et~al.}{2024}]{sajun_historical_2024}
Sajun A.~R.,  Zualkernan I.,   Sankalpa D.,  2024, \mn@doi [Applied Sciences] {10.3390/app14104316}, 14, 4316

\bibitem[\protect\citeauthoryear{Shen, Li, Chen, Yan, Quan, Chen, Zhang  \& Huang}{Shen et~al.}{2024}]{shen_small_2024}
Shen W.,  Li C.,  Chen H.,  Yan M.,  Quan X.,  Chen H.,  Zhang J.,   Huang F.,  2024, Small {LLMs} {Are} {Weak} {Tool} {Learners}: {A} {Multi}-{LLM} {Agent}, \mn@doi{10.48550/arXiv.2401.07324}, \url {https://ui.adsabs.harvard.edu/abs/2024arXiv240107324S}

\bibitem[\protect\citeauthoryear{Souly et~al.,}{Souly et~al.}{2025}]{souly_poisoning_2025}
Souly A.,  et~al., 2025, Poisoning {Attacks} on {LLMs} {Require} a {Near}-constant {Number} of {Poison} {Samples}, \mn@doi{10.48550/arXiv.2510.07192}, \url {http://arxiv.org/abs/2510.07192}

\bibitem[\protect\citeauthoryear{Tamhane}{Tamhane}{2025}]{tamhane_natural_2025}
Tamhane P.,  2025, A {Natural} {Language} {Interface} for {Efficient} {Data} {Retrieval} in {SDSS}, \mn@doi{10.48550/arXiv.2510.25953}, \url {https://ui.adsabs.harvard.edu/abs/2025arXiv251025953T}

\bibitem[\protect\citeauthoryear{Taylor}{Taylor}{2005}]{taylor_topcat_2005}
Taylor M.~B.,  2005. p.~29, \url {https://ui.adsabs.harvard.edu/abs/2005ASPC..347...29T}

\bibitem[\protect\citeauthoryear{Taylor}{Taylor}{2006}]{taylor_stilts_2006}
Taylor M.~B.,  2006. p.~666, \url {https://ui.adsabs.harvard.edu/abs/2006ASPC..351..666T}

\bibitem[\protect\citeauthoryear{Vaswani, Shazeer, Parmar, Uszkoreit, Jones, Gomez, Kaiser  \& Polosukhin}{Vaswani et~al.}{2017}]{vaswani_attention_2017}
Vaswani A.,  Shazeer N.,  Parmar N.,  Uszkoreit J.,  Jones L.,  Gomez A.~N.,  Kaiser L.,   Polosukhin I.,  2017, Attention {Is} {All} {You} {Need}, \mn@doi{10.48550/arXiv.1706.03762}, \url {https://ui.adsabs.harvard.edu/abs/2017arXiv170603762V}

\bibitem[\protect\citeauthoryear{Wolf et~al.,}{Wolf et~al.}{2020}]{wolf_transformers_2020}
Wolf T.,  et~al., 2020, in Liu Q.,  Schlangen D.,  eds, Proceedings of the 2020 {Conference} on {Empirical} {Methods} in {Natural} {Language} {Processing}: {System} {Demonstrations}. Association for Computational Linguistics, Online, pp 38--45, \mn@doi{10.18653/v1/2020.emnlp-demos.6}, \url {https://aclanthology.org/2020.emnlp-demos.6/}

\makeatother
\end{thebibliography}

\appendix

\section{Evaluation similarity score by task}
\begin{figure*}
    \centering
    \includegraphics[width=\linewidth]{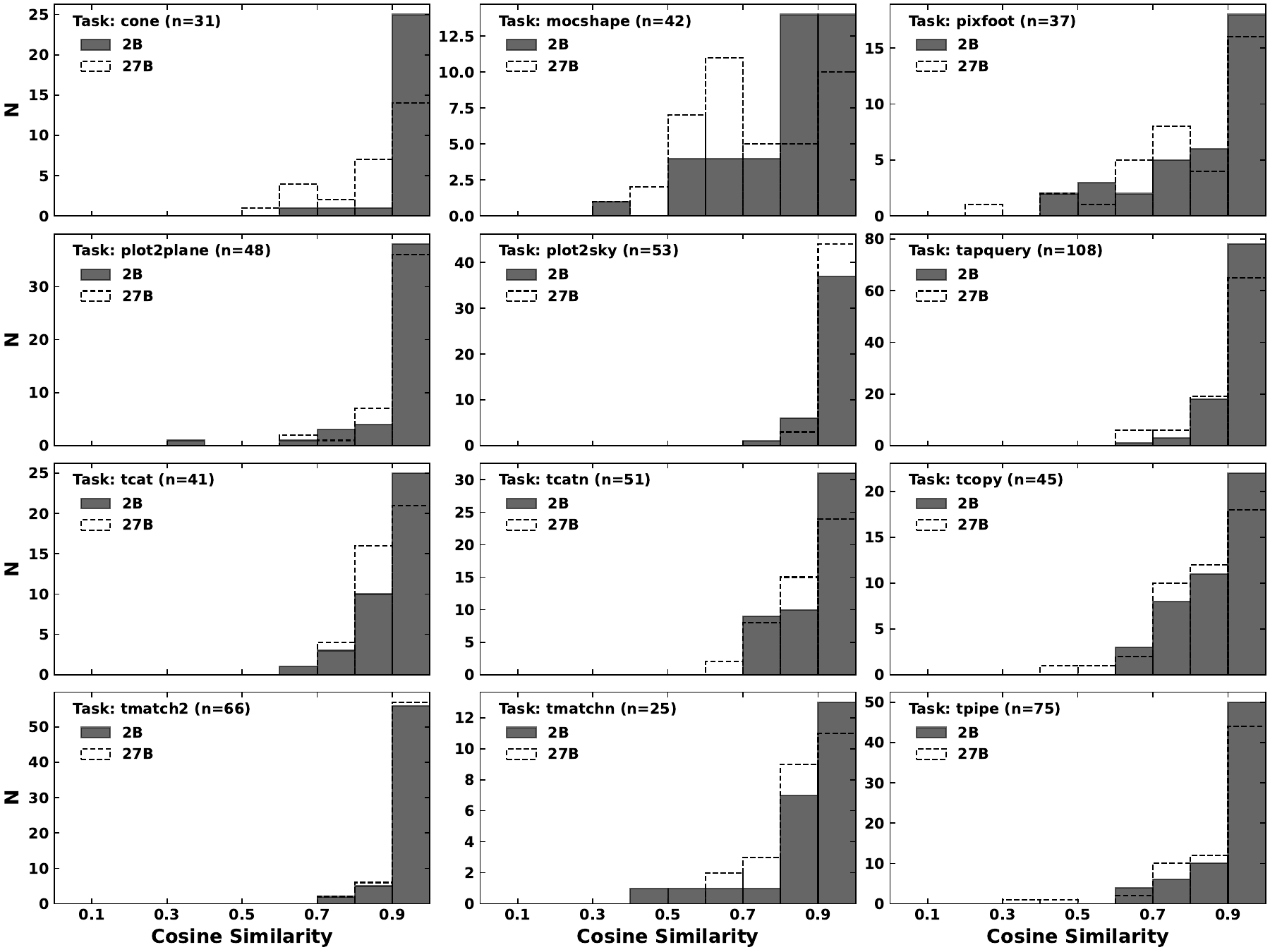}
    \caption{Cosine similarity histogram separated by STILTS task.}
    \label{fig:sim_hist_by_task}
\end{figure*}

\section{Example user interaction with \NAME\ }
\label{appen:example_user_interaction}

\begin{figure}
\begin{verbatim}
Welcome to STILTS-NLI!
>> Hello!

Hello! It's nice to meet you. How can I assist you 
today?

>> Can you make me a STILTS command that will replace 
all null values in 'demo_cat.fits' column MAG to -99.

stilts tpipe in=demo_cat.fits cmd='replaceval null -99 
MAG' out=demo_cat_nulls_removed.fits

>> Now can you execute this on my machine?

I can help you with executing the STILTS command 
on your machine.

stilts tpipe in=demo_cat.fits cmd='replaceval null -99 
MAG' out=demo_cat_nulls_removed.fits

Do you want to execute this? (y/n): y

Command executed successfully.
\end{verbatim}
\caption{Example interaction with STILTS-NLI showing both STILTS general greeting, command generation and execution on the user's machine.}
\end{figure}

\begin{figure}
\begin{verbatim}
    
Welcome to STILTS-NLI!

>> I want some stats on my catalogue 'demo_cat_fixed.fits'. 
Can you make a command for this?

stilts tpipe in=demo_cat_fixed.fits cmd='keepcols "MAG"' 
omode=stats

>> Please run this.

stilts tpipe in=demo_cat_fixed.fits cmd='keepcols "MAG"' 
omode=stats

Do you want to execute this? (y/n): y

Command executed successfully.
Output:
Total Rows: 10
+--------+------------+----------+-------+-----------+------+
| column | mean       | stdDev   | min   | max       | good |
+--------+------------+----------+-------+-----------+------+
| MAG    | -23.052374 | 49.89358 | -99.0 | 19.126186 | 10   |
+--------+------------+----------+-------+-----------+------+

\end{verbatim}
\caption{Example interaction with STILTS-NLI showing STILTS command generation and execution with STILTS output.}
\end{figure}
\bsp	
\label{lastpage}
\end{document}